\documentclass[11pt]{article} 
\usepackage[a4paper, margin=1.25in]{geometry} 
\usepackage[T1]{fontenc}
\usepackage[normalem]{ulem}

\usepackage[ruled,vlined]{algorithm2e}
\usepackage{graphicx}
\usepackage{booktabs}
\usepackage{xcolor}
\usepackage{rotating}

\newcommand{\para}[1]{\textbf{#1.}}

\usepackage[dvipsnames,svgnames,x11names]{xcolor}

\colorlet{RED}{red}
\colorlet{WHITE}{white}
\colorlet{FORESTGREEN}{ForestGreen}
\colorlet{MAROON}{Maroon}

\usepackage[nocompress]{cite}
\usepackage{balance}
\usepackage{listings}

\lstdefinelanguage{XML}
{
basicstyle=\ttfamily\footnotesize,
  morestring=[b]",
  moredelim=[s][\bfseries\color{Maroon}]{<}{\ },
  moredelim=[s][\bfseries\color{Maroon}]{</}{>},
  moredelim=[l][\bfseries\color{Maroon}]{/>},
  moredelim=[l][\bfseries\color{Maroon}]{>},
  morecomment=[s]{<?}{?>},
  morecomment=[s]{<!--}{-->},
  commentstyle=\color{gray},
  stringstyle=\color{blue},
  identifierstyle=\color{red}
}

\usepackage{moreverb}
\usepackage[nounderscore]{syntax}
\graphicspath{{./arXiv-version/arXiv-figures}}
\DeclareGraphicsExtensions{.pdf}
\usepackage[cmex10]{amsmath}
\usepackage{amssymb}
\usepackage{mathtools}
\usepackage{amsthm}
\usepackage{amsfonts}

\usepackage{subfig}

\usepackage{multirow}
\usepackage{rotating}
\usepackage{booktabs}
\usepackage{colortbl}
\usepackage{tablefootnote}

\usepackage{array}
\newcolumntype{L}[1]{>{\raggedright\let\newline\\\arraybackslash\hspace{0pt}}m{#1}}
\newcolumntype{C}[1]{>{\centering\let\newline\\\arraybackslash\hspace{0pt}}m{#1}}
\newcolumntype{R}[1]{>{\raggedleft\let\newline\\\arraybackslash\hspace{0pt}}m{#1}}

\usepackage[pdftex,colorlinks=true,urlcolor=blue,citecolor=blue]{hyperref}
\usepackage{xspace}

\usepackage{enumitem}

\usepackage[utf8]{inputenc}
\usepackage[T1]{fontenc}
\usepackage{url}
\usepackage{booktabs}
\usepackage{amsfonts}
\usepackage{nicefrac}
\usepackage{lmodern}
\usepackage{microtype}
\usepackage{lipsum}
\usepackage{rotating}
\usepackage{multirow}

\begin{document}

\title{Scaling Real-Time Traffic Analytics on Edge--Cloud Fabrics for City-Scale Camera Networks \footnote{Accepted at TCSC SCALE Challenge 2026. To appear in the Proceedings of IEEE/ACM CCGRID Workshops, Sydney, 2026.}}
\author{
    Akash Sharma$^{1}$, Pranjal Naman$^{1}$, Roopkatha Banerjee$^{1}$,
    Priyanshu Pansari$^{1}$,\\ Sankalp Gawali$^{1}$, Mayank Arya$^{1}$, 
    Sharath Chandra$^{1}$, Arun Josephraj$^{3}$,\\ Rakshit Ramesh$^{3}$, Punit Rathore$^{2,4}$
    Anirban Chakraborty$^{1}$,\\ Raghu Krishnapuram$^{3,4}$,
    Vijay Kovvali$^{4}$ and Yogesh Simmhan$^{1,2}$\\~\\
    \small $^{1}$Department of Computational and Data Sciences\\
    \small $^{2}$Robert Bosch Center for Cyber Physical Systems (RBCCPS)\\
    \small $^{3}$Center of Data for Public Good (CDPG)\\
    \small $^{4}$Centre for infrastructure, Sustainable Transportation \& Urban Planning\\
    \em Indian Institute of Science, Bengaluru, India\\~\\
    \texttt{Email: \{akashsharma, pranjalnaman, simmhan\}@iisc.ac.in}
}

\date{}
\maketitle
\begin{abstract}

Real-time city-scale traffic analytics requires processing 100s--1000s of CCTV streams under strict latency, bandwidth, and compute limits. We present a scalable AI-driven Intelligent Transportation System (AIITS) designed to address multi-dimensional scaling on an edge--cloud fabric. Our platform transforms live multi-camera video feeds into a dynamic traffic graph through a DNN inferencing pipeline, complemented by real-time nowcasting and short-horizon forecasting using Spatio-Temporal GNNs. Using a testbed to validate in a Bengaluru neighborhood, we ingest 100+ RTSP feeds from Raspberry Pis, while Jetson Orin edge accelerators perform high-throughput detection and tracking, producing lightweight flow summaries for cloud-based GNN inference. A capacity-aware scheduler orchestrates load-balancing across heterogeneous devices to sustain real-time performance as stream counts increase. To ensure continuous adaptation, we integrate SAM3 foundation-model assisted labeling and Continuous Federated Learning to update DNN detectors on the edge. Experiments show stable ingestion up to 2000 FPS on Jetson Orins, low-latency aggregation, and accurate and scalable ST-GNN forecasts for up to 1000 streams. A planned live demonstration will scale the full pipeline to 1000 streams, showcasing practical, cross-fabric scalability.
\end{abstract}

\section{Introduction}
\para{Social Impact of Traffic Congestion}
Traffic congestion and safety remain critical challenges in rapidly urbanizing regions, impacting quality of life, public health, and energy sustainability. This is especially severe in emerging megacities with >10M populations. Bengaluru, India, the second most congested city globally, experiences average commute speeds of 14~kmph (8.7~mph), leading to nearly $7$ lost days annually per commuter~\cite{tomtom-2025}.
Road transport contributes $\approx 30\%$ of global energy use, making congestion a sustainability concern~\cite{iea_energy_efficiency_2025_transport}.
Air quality also degrades with traffic density~\cite{efd_dp_23_10_congestion_health_hazards}.
India reports the world's highest road fatalities, underscoring the need for systemic interventions~\cite{grsf_annual_report_2024}.

\para{An AI-Driven Approach to ITS}
These realities motivate the need for \textit{Intelligent Transportation Systems (ITS)} that can operate effectively under heterogeneous, unstructured, and dynamic traffic conditions.
Unlike lane-disciplined traffic in developed regions, Indian roads exhibit mixed vehicle types and models, ranging from two-wheelers and three-wheelers to buses and trucks, moving in non-lane-based patterns. 
Compounding this, developing cities are also under-invested in specialized sensing infrastructure for traffic monitoring, such as lane or ANPR (Automatic Number Plate Recognition) cameras, relying instead on thousands of general-purpose safety cameras that were not designed for traffic flow analytics.
As a result, traditional computational traffic simulators (e.g., SUMO, Vissim) cannot be calibrated effectively for these conditions to optimize traffic \cite{naing2024fine}.
These constraints motivate \textit{AI-driven ITS (AIITS)} that infer traffic states from large, heterogeneous camera networks.

\para{Need for Speed in AIITS}
AIITS pipelines are computationally and communication-intensive.  Bengaluru, our case-study city, has $>5000$ safety cameras that generate continuous HD video feeds. 
Today's centralized architectures rely on high-bandwidth fiber networks, GPU servers, and petabytes of archival storage for even \textit{manual monitoring}.
Transitioning to AI-driven ITS solutions requires \textit{active real-time computation}, not just \textit{passive storage} over the streams.
Scaling to 100s of gigabits and 100k+ Frames per Second (FPS) of video streams implies unsustainable growth in cloud GPU servers. Further, actionable traffic forecasting requires low-latency ingestion and inference, which centralized systems struggle to provide across Wide Area Networks (WANs) as camera deployments increase.
This defines our problem: \textit{enabling real-time analytics from 1000s of city cameras under strict latency and resource constraints, requiring coordinated scale-out, scale-up and elastic stream distribution.}

\para{Contributions}
We introduced a scalable \textit{edge--cloud framework} for real-time AI-driven traffic analytics over city-wide camera networks. Our system processes 100s of heterogeneous video streams under strict latency, bandwidth, and compute constraints by designing an \textit{edge-first design} that performs high-throughput detection and tracking on distributed Jetson accelerators, while Raspberry Pis serve large-scale RTSP video streams. 
The framework demonstrates \textit{multi-dimensional scalability}: (i) \textit{scale-out} across 50+ heterogeneous edge devices and 100 streams, (ii) \textit{scale-up} of GNN pipelines to sustain real-time inference of up to 1000 streams, and (iii) \textit{cross-fabric scaling} from edge to cloud. 

A \textit{capacity-aware stream scheduler} intelligently assigns video streams to edge accelerators based on empirical FPS thresholds, to sustain real-time performance as streams scale from tens to hundreds. Structured per-camera traffic flow summaries are fused into a dynamic city-level temporal traffic graph feeding a \textit{GNN-based forecasting} service for short-horizon congestion predictions. To maintain analytical freshness, a \textit{Continuous Federated Learning} pipeline detects \textit{de novo} vehicle classes using a \textit{foundation model} and retrains DNN detectors on the edge, concurrently with inference. 

Our end-to-end evaluation for a \textit{neighborhood in Bengaluru} with 100 cameras shows sustained real-time throughput, low latency, and robust scaling. This demonstrates a deployable and socially impactful AIITS platform. We plan to expand this to 1000 streams in a live demonstration.

\section{Towards AI-driven ITS in Bengaluru}
\begin{figure}[!t]
  \centering
	\includegraphics[width=0.8\columnwidth]{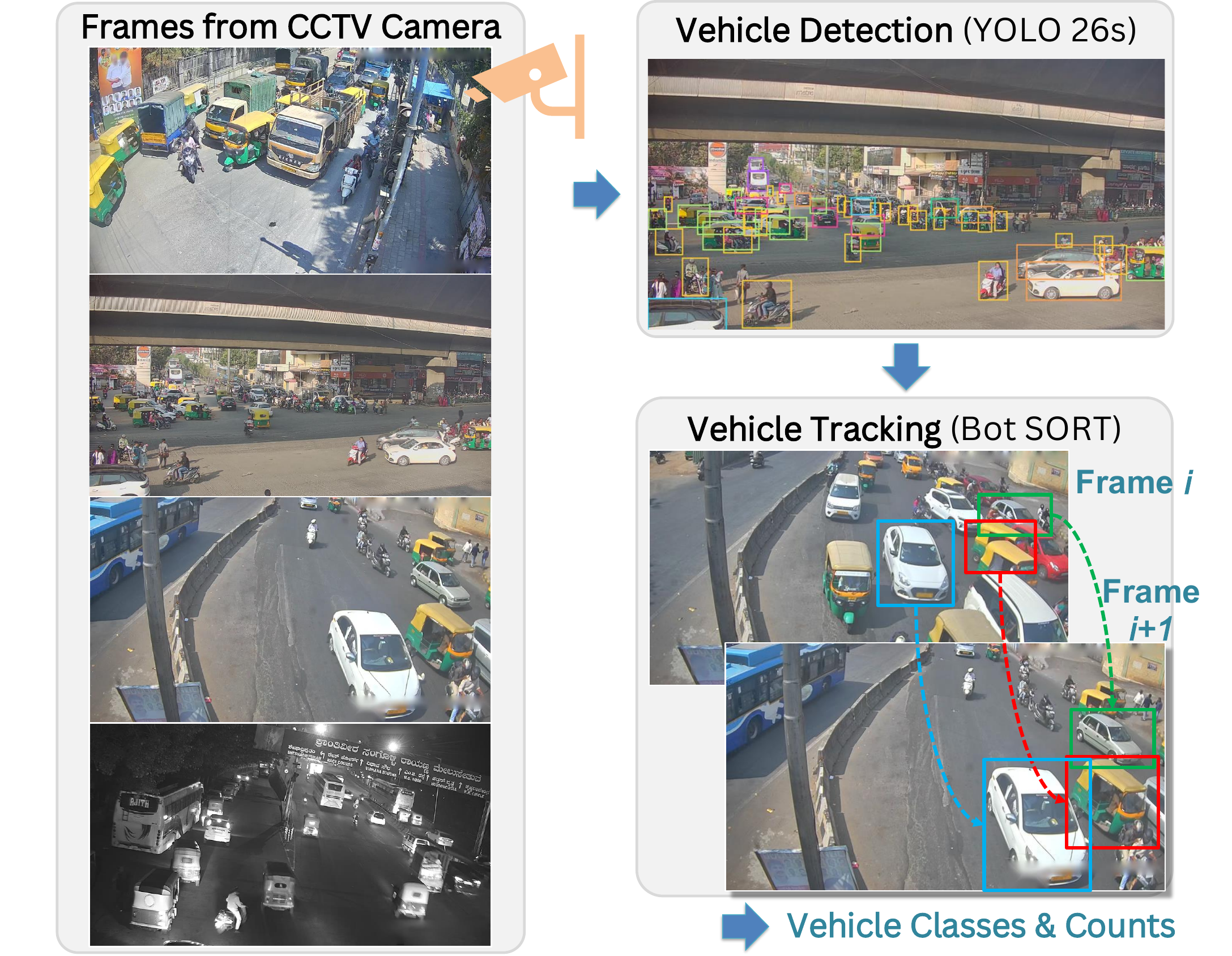}
	\caption{Sample images reflecting complex traffic in Bengaluru, with a DNN pipeline to detect vehicle classes and counts.}
    \label{fig:video}
    \vspace{-0.1in}
\end{figure}

The \textit{AI for Mobility at IISc (AIM@IISc)} is an inter-disciplinary team that develops scalable, data-driven tools to improve congestion and safety on Indian roads. Bengaluru serves as an ideal real-world testbed, with close engagement from the city traffic police to design and validate novel and scalable methods.
We introduce the high-level AIITS requirements here, followed by the proposed design in \S~\ref{sec:arch}.

A critical requirement for scalable AIITS is converting 5000+ safety camera streams into a \textit{temporal graph} that exposes real-time network conditions. This transformation reduces the high-volume, bandwidth-heavy multi-dimensional data 
into a concise temporal graph form suitable for downstream analytics. We address this through an edge-first inferencing pipeline (Fig.~\ref{fig:video}) coupled with cloud-based aggregation.

Multiple cameras predominantly placed at road junctions cover only a fragment of the 2000+~kms road network across a 700~km$^2$ region. This requires \textit{spatial extrapolation} from the $\approx 1000$ observed vertices and edges to the 10k+ unobserved roads across the city. Given the limitations of calibrating classical traffic simulators for India's heterogeneous traffic, we adopt \textit{Spatio-temporal Graph Neural Network (ST-GNN)},
enabling both real-time \textit{nowcasting} and \textit{short-term forecasting} of traffic across the broader network.

These forecasts support higher-level analytics that help inform interventions. 

\textit{Anomaly detection} over the predicted temporal graph can flag segments with unusually high congestion, supporting targeted traffic police deployment or remote signal control. Such wide-area, real-time signal optimization can be increasingly automated.  The ST-GNN models also support \textit{what-if analysis}, allowing planners to evaluate policy options such as introducing one-way flows, adjusting lane ratios or adding bus-only lanes. These reduce the cost of the solution and promote evidence-driven urban mobility decisions.

Finally, training accurate vision models for Indian traffic is challenging due to heterogeneous vehicle types and evolving conditions. Our prior large-scale annotation effort (e.g., UVH-26 ~\cite{sharma2025uvh26}) improved the accuracy of state-of-the-art YOLO and DETR models from $8.4$ to $31.5$\% over equivalent baseline models trained on the COCO dataset \cite{coco}, but remains human-intensive and cannot scale across cities. This motivates using \textit{multi-modal foundational models}, e.g., SAM3, CLIP, as automatic teachers to bootstrap localized training data. We integrate this into a \textit{Continuous Federated Learning} setup that enables scalable, distributed retraining on edge accelerators, without interrupting real-time inference. This demonstrates scalable application--infrastructure co-design.

\section{Scalable AIITS Architecture and Testbed}
\label{sec:arch}

\begin{figure*}[!t]
\vspace{-0.1in}
  \centering
	\includegraphics[width=1\textwidth]{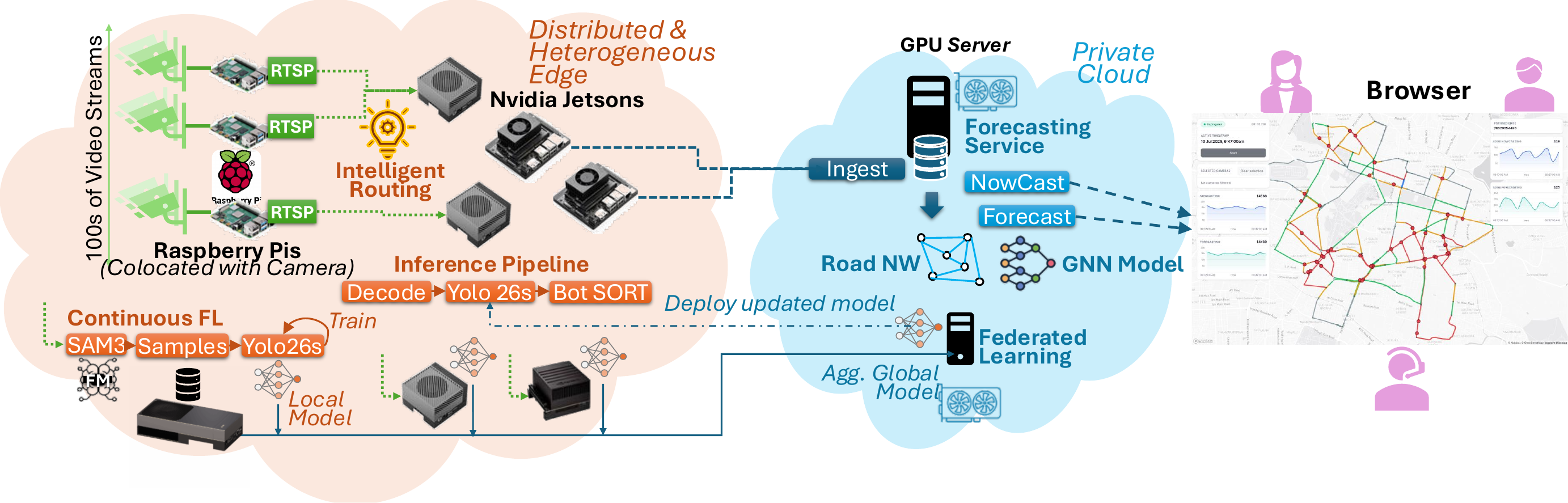}
	\caption{Scalable AIITS Testbed and Architecture for Bengaluru Traffic Camera Feeds.} 
    \label{fig:arch}
    \vspace{-0.1in}
\end{figure*}

Fig.~\ref{fig:arch} illustrates the architecture of our AI-driven ITS platform, which spans edge and cloud resources.
We instantiate an \textit{AIITS testbed} that emulates the proposed city-scale deployment at an $\mathbb{O}(100)$ scale stream. Each video camera has an embedded computer that hosts an RTSP server to serve the streams, proxied in our testbed using Raspberry Pis. The core computational fabric consists of  \textit{Nvidia Jetson Edge accelerators}, including Orin AGX and Thor developer kits, which provide multi-core Arm CPUs, 1792--2560 Ampere/Blackwell CUDA cores and 32--128GB of share CPU--GPU LPDDR memory in compact (11cm$\times$11cm$\times$7.2cm, 800--1800~gms) low-powered (60--130W peak) form factor for field deployment. These devices are colocated with cameras at certain roadside units and ingest video streams over the local fiber network to perform real-time inferencing and training.

A small number of centralized GPU workstations, hosted at the City's private cloud, orchestrate the platform, ingest light-weight structured flow updates, maintain the temporal traffic graph, execute nowcasting and forecasting using GNN models, and manage federated learning aggregation. These also host browser-based services that expose live forecasts, alerts, and system controls to city operators, residents, and other stakeholders. We discuss these various components next.

\subsection{Streaming RTSP Streams from Edge}
To evaluate our system at scale without relying on physical CCTV deployments, we construct a distributed testbed that emulates a large urban camera network using hundreds of concurrent \emph{Real-Time Streaming Protocol (RTSP)} video streams. Each stream is hosted on a heterogeneous cluster of Raspberry Pis, where every Pi runs \textit{MediaMTX}, an open-source RTSP server. Each server exposes one or more streams through unique endpoints for downstream processing.
Video is served using \textit{FFmpeg}, which replays pre-recorded camera footage stored locally on each Pi for specific time durations. 
By using FFmpeg's \textit{stream-copy}, we avoid transcoding overheads, enabling multiple real-time streams even on low-power Pis. This architecture allows high concurrency across the cluster. Collectively, our 42 Pis of 3B and 4B types are able to serve 100 distributed streams, forming a scalable, low-cost emulation of a city-wide camera network neighborhood. This also weakly-scales with the addition of more Pis to support a larger number of streams.

\subsection{Scaling Detection Models on Edge Accelerators}
To convert raw multi-camera video streams into data suitable for GNN-based forecasting, we deploy a distributed edge inference layer across heterogeneous Nvidia Jetson devices in our testbed. This layer ingests live RTSP streams, performs real-time detection and tracking, and produces compact vehicle-flow summaries for downstream analytics. By executing inference locally, the system avoids high-bandwidth video transmission to the cloud and reduces end-to-end latency.

\subsubsection{Stream Ingestion} 
End-users select camera streams via a frontend interface, which are then dynamically assigned to Jetson accelerators (\S~\ref{sec:intelligent-routing}) by our scheduler. 
On each Jetson~(Fig.~\ref{fig:arch}), hardware-accelerated \textit{GStreamer} pipelines utilize the on-board \textit{nvdec} accelerator to decode streams directly into GPU-native memory. 
DeepStream's \textit{nvstreammux} batches frames from multiple camera streams into unified inference batches, enabling high GPU utilization and allowing Jetsons to process many simultaneous streams at real-time frame rates.

\subsubsection{Detection and Tracking} 
Each batched frame is processed using the YOLO26s model~\cite{ultralytics_yolo}, trained on our UVH-26 dataset~\cite{sharma2025uvh26} and deployed using DeepStream's inference engine. While object detection provides per-frame bounding boxes and classes, real-time traffic estimation requires tracking \textit{unique vehicles} in a feed. 
To accomplish this, each stream integrates BoT-SORT multi-object tracking~\cite{aharon2022botsortrobustassociationsmultipedestrian}, which assigns persistent IDs to detected vehicles across successive frames. For every processed frame, the system emits vehicle-level records consisting of a tracking ID, class label, and bounding box. These identity-aware observations are aggregated over short intervals by our pipeline (tunable, 5--30s) and forwarded as low-overhead summaries to the GNN inference pipeline, avoiding the need to transmit raw video or dense feature maps.

\subsubsection{Intelligent Routing} \label{sec:intelligent-routing} 
Because the inference layer operates over heterogeneous Jetson devices, effective stream placement is essential for maintaining real-time performance. 
We enable elastic scaling by assigning streams to accelerators using a \textit{capacity-driven scheduling policy},
based on empirically derived per-device limits. For each Jetson model, we profile it offline by progressively increasing the input workload until the device no longer sustains the native frame rate, defining its stable throughput. In our deployment, Orin AGX 32GB and 64GB models support $\approx 200$~FPS and $400$~FPS, respectively. Stream assignment is then formulated as a bin-packing problem~\cite{coffman1984approximation}, where each stream's FPS contributes to the item weight and each device represents a bin with known capacity. We evaluate two classic heuristics:

\paragraph{Best Fit}
Streams are placed on the smallest-capacity device that can still accommodate them. In our testbed devices (\S~\ref{sec:setup}), this packs the loads tightly onto the Orin 32GB devices first, and activates the larger Orin 64GB devices only when cumulative demand exceeds $\approx1000$ FPS. This minimizes the number of active devices and reduces overall power consumption at moderate workloads.

\paragraph{Worst Fit}

Streams are assigned to the device with the largest remaining capacity, distributing the load across even the higher-capacity devices earlier. This improves load balance and thermal distribution (e.g., when it is a hot day and the IP67 packaging imposes thermal limits) but activates more devices sooner, increasing baseline cluster power usage.

These strategies allow the system to flexibly scale detection and tracking across heterogeneous edge accelerators while accounting for power usage and sustaining real-time throughput for large, multi-camera deployments.

\subsection{Traffic Forecasting on Temporal Graphs using GNNs}

The GPU server hosts the \textit{ingest, nowcast} and \textit{forecast services} that operate on the structured vehicle-flow outputs received from the edge inference layer~(Fig.~\ref{fig:arch}). 
This represents the scale-up aspect of our architecture, leveraging GPU servers for high-throughput graph inference.
A local GPU workstation in our testbed mimics an accelerated VM in the private cloud. 
The \textit{ingest service} receives per-camera vehicle counts at 1-second granularity, batched every 15 seconds, and maintains an on-disk time-series database with timestamp, camera ID, and a class-count vector. The nowcast service exposes the aggregated traffic state to downstream applications through a low-latency streaming interface implemented as a gRPC service.

The forecast service applies a \textit{Spatio-Temporal Graph Neural Network (ST-GNN)} to predict short-horizon traffic conditions across the road network. ST-GNNs jointly model the spatial structure of road graphs and the temporal evolution of traffic patterns. We use the TrendGCN architecture~\cite{TrendGCN}, which integrates vertex embeddings, temporal embeddings and a graph-convolutional GRU to capture dynamic traffic correlations across junctions. 
It is trained offline by aggregating tracker outputs into 1-minute junction-level vehicle-count time series and using 180 hours of historical data from 100 junctions with a lag of 5 minutes and a 5-minute forecast horizon. The model queries historical data from the ingest database based on a configurable lag window (e.g., several minutes), and outputs predicted vehicle counts for each junction over the chosen forecast horizon.

Because camera coverage is incomplete, only $\approx100$ of the over 250 junctions are observed in our validation neighborhood; we forecast on a coarsened graph that retains only camera-equipped junctions as nodes. Edges in this reduced graph are super-edges formed by collapsing sequences of road segments between observed junctions, preserving large-scale structure for short-range traffic propagation while ensuring reliable inputs at every node~\cite{li2018diffusion}. To obtain street-level (edge) flow estimates from predicted junction-level counts, we apply a mass-conserving allocation that distributes vehicle volumes across adjacent super-edges proportional to their connectivity, and aggregate endpoint contributions to compute edge counts, maintaining consistency between vertex and edge predictions for congestion estimation.

Finally, edge-level flows are discretized into congestion states (e.g., free-flow, moderate, heavy) for dashboard integration and downstream analytics. By combining high-frequency edge summaries, spatio-temporal forecasting, and graph-based abstractions, the system delivers accurate short-horizon predictions that scale to city-wide deployments under partial camera coverage.

\subsection{LLM-driven Labeling for Federated Learning on Edge}
\label{subsec:fl}

Edge-based detection models must continuously adapt to highly localized and evolving vehicle categories, such as auto-rickshaws (3-wheelers) or emerging commercial vehicle types, that are not well represented in standard datasets. To support this, we implement a continuous FL workflow across $K$ Jetson devices, enabling each edge device to curate, annotate, and train on its own data without transmitting raw video to a central server.
Each Jetson collects training samples through temporally stratified sampling.
Frames are accumulated over a configurable window to form a local, diverse dataset.

After the collection phase, each device uses the SAM3 foundation model~\cite{carion2025sam} to automatically annotate the sampled images.
The annotation pipeline accepts an image and a predefined set of text prompts, e.g., 

$C = \{ \text{``\textit{a sedan}''}, \text{``\textit{a sport-utility vehicle}''}, \cdots \}$,
which cover common and region-specific vehicle classes, and a subset of which may be unknown to the object detection model. 
These prompts are tokenized into embeddings $E_c$ and the SAM3 model $f_\theta$ processes these inputs to 
generate bounding box coordinates $B_c$ and raw class logits $Z_c$ for a batch of data $P$:$(Z_c, B_c) = f_\theta(P; E_c)$.
Logits are converted to probabilities using a sigmoid function, and only predictions exceeding a confidence threshold $\tau$ (e.g., $\tau=0.30$) are retained. The resulting per-device dataset is
$D_k = \bigcup_{c \in \mathcal{C}} \{ (c, \text{bbox}_q, p_q) \mid p_q(c) \ge \tau \}$.

Each Jetson then performs local fine-tuning of the YOLO-based detector, initialized with the current global weights. After $E$  local epochs of training over $D_k$, devices return their updated weights to the server, which aggregates them using Federated Averaging (FedAvg).
This aggregated global model is then broadcast back to the edge devices for use in the detection inference pipeline.

\subsection{Dashboard}

We implement an interactive React+Mapbox dashboard (Fig.~\ref{fig:arch}, right) as the system’s primary visualization and control interface. Outputs are overlaid on an OSM map of Bengaluru, with camera locations and inferred road segments rendered geospatially. Updates stream via gRPC to a FastAPI backend and are forwarded to the browser over WebSockets to maintain low-latency, continuous refresh. The map provides a live nowcast view, showing how per-camera vehicle-flow summaries translate into city-level traffic conditions. Color-coded overlays indicate congestion levels, enabling operators to assess the network state at a glance. A complementary time-series panel visualizes real-time and forecasted vehicle counts, updated as the ingestion and forecasting services produce new outputs. Displaying short-horizon GNN forecasts alongside current measurements helps highlight emerging congestion trends.

The dashboard also supports lightweight orchestration. Users can select individual cameras or subsets of streams to trigger targeted execution of the detection and tracking pipeline. Selecting a road segment opens a historical summary of vehicle flow, derived from second-wise aggregated counts maintained by the ingest service. 

\section{Results}
\label{sec:results}

\subsection{Setup}
\label{sec:setup}

Our RPi cluster hosts 1000 camera streams, with 10 Pi 4B/8GB with 4 streams each,
17 Pi 4B/2GB with 2--33 streams each, 
, and 15 Pi 3B/1GB with 1 stream. 
Video feeds are usually at 25 FPS, pre-synchronized across Pis
and streamed by the RTSP servers.
The testbed also has an accelerated edge cluster Jetson Orin AGX, five with 32GB and 4 with 64GB, to host the inference pipelines, and ingest these 100 video streams. The FL setup runs on these same devices

\subsection{Scalability of Edge Analytics for Vehicle Counting}

\begin{figure}[!t]
\vspace{-0.1in}
  \centering
    \includegraphics[width=\columnwidth]{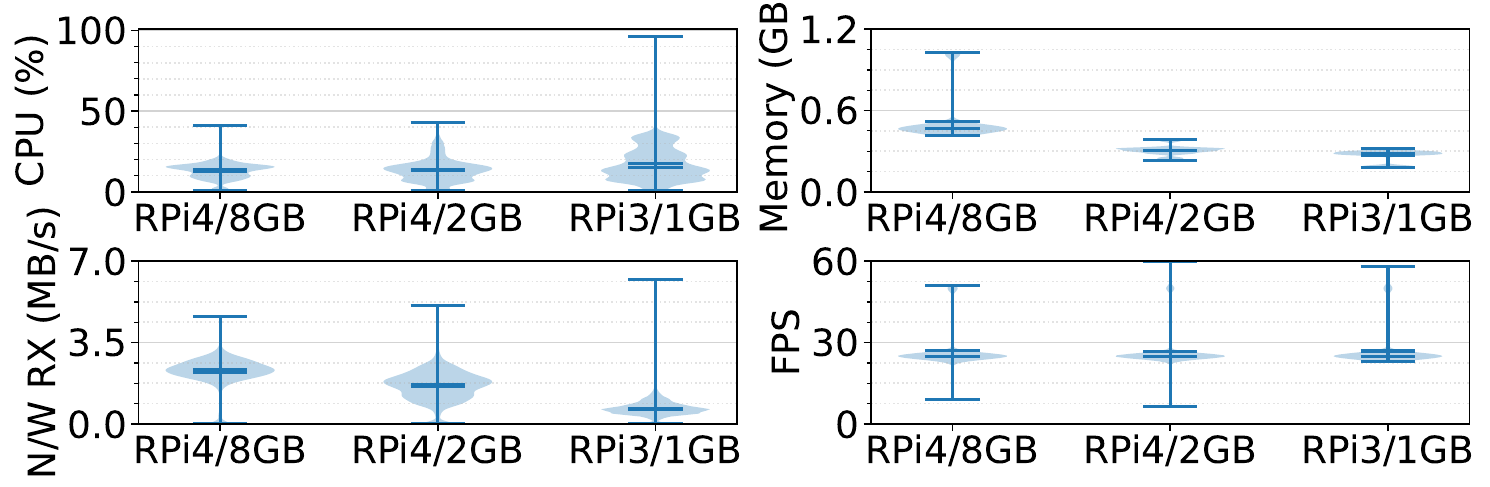}
	\caption{Performance of RPis hosting 100 RTSP Streams}
    \label{fig:results_rpi}
\end{figure}

\subsubsection{RPi RTSP Stream Performance}
To validate testbed stability, we evaluate resource usage across Raspberry Pi variants (Fig.~\ref{fig:results_rpi}). The results show that overhead on these RTSP publisher devices is minimal: median CPU utilization stays below 25\% on all models, largely due to the stream-copy mechanism avoiding costly encode/decode operations. Memory usage remains modest as well, peaking at 30\% on the RPi 3/1GB, with the RPi4/8GB slightly higher because it hosts four streams. Network throughput follows stream count: RPi 4/8GB sends the most data, while RPi 4/2GB send two or three streams, but all remain within bandwidth limits at $\leq 7$~MB/s, below the 100Mbps or $12.5$~MB/s cap of RPi3. 
FPS remains stable, within $25\pm1$FPS, $>90\%$ of the time.
These results confirm that even low-power Pis can reliably scale to high stream densities for our AIITS testbed without degrading stream quality.

\begin{figure}[!t]
\vspace{-0.2in}
  \centering
  \subfloat[FPS Capacity Use\% and Cumulative FPS Processed]{
    \includegraphics[width=0.9\linewidth]{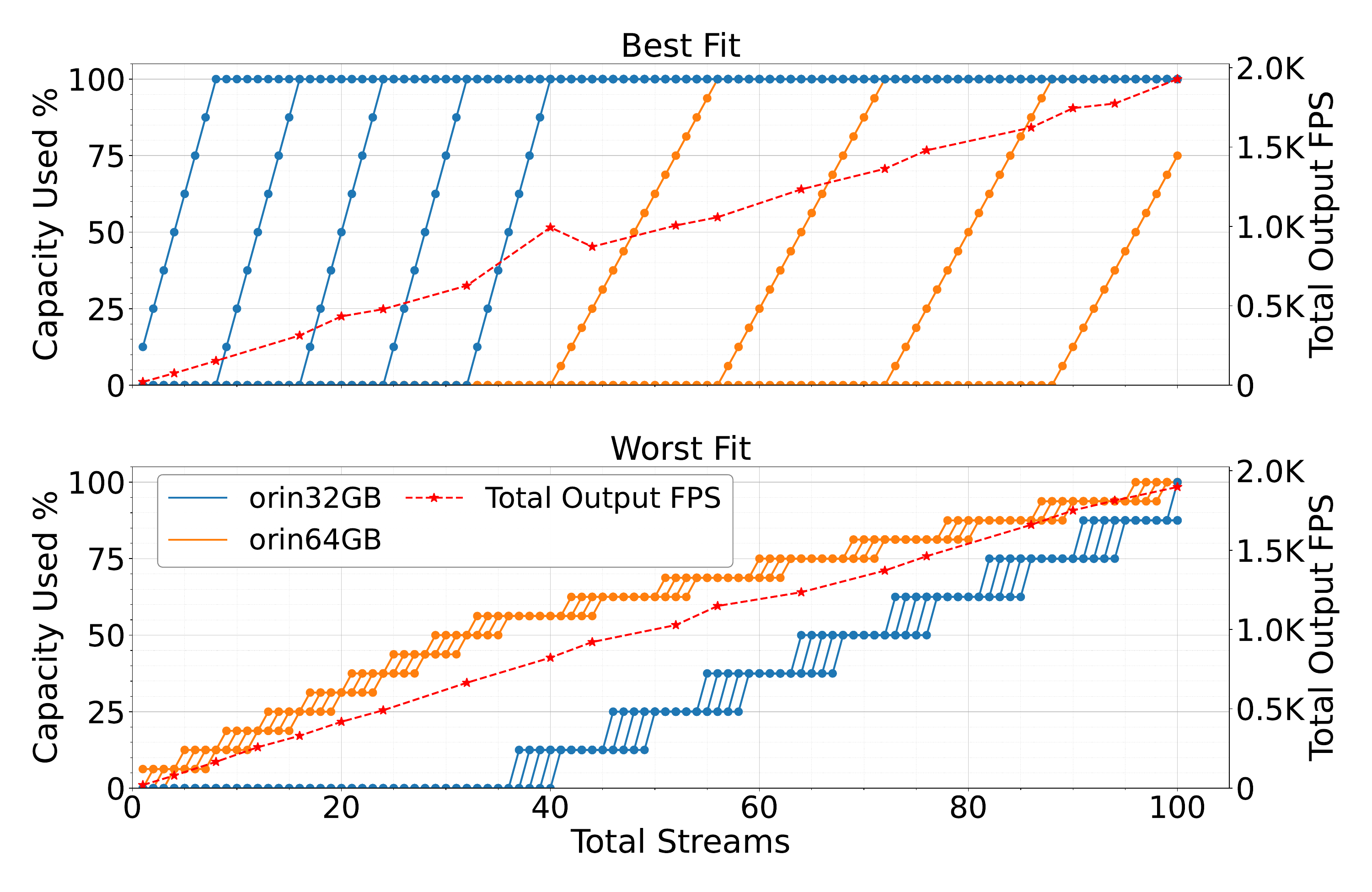}
    \label{fig:edge_results_1}
  }\\
  \vspace{-0.1in}
  \subfloat[Active and Used TFLOPS\% in cluster, and Total Power Draw]{
    \includegraphics[width=0.9\linewidth]{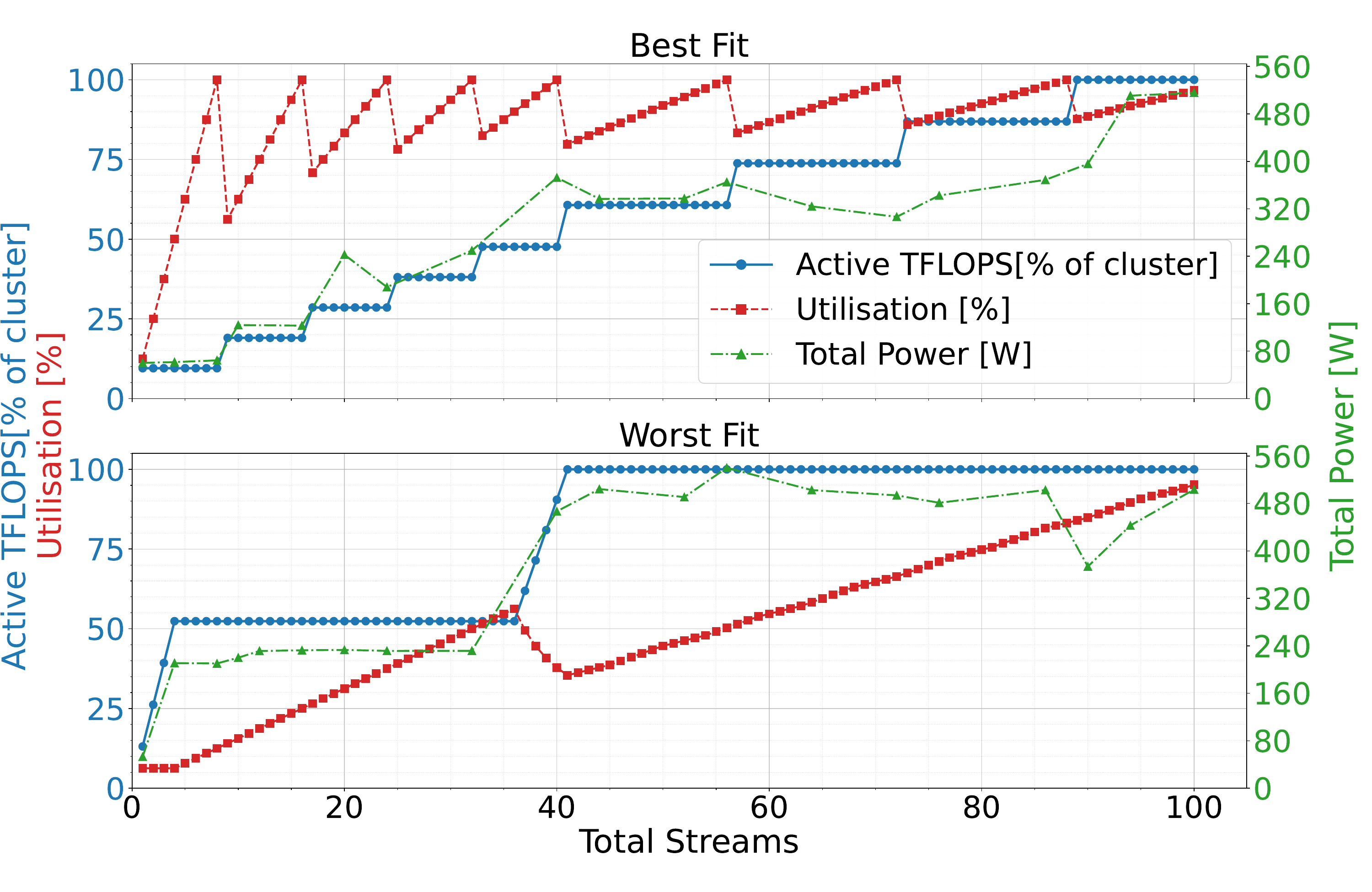}
    \label{fig:edge_results_2}
  }
  \caption{Efficiency and Scaling of Jetsons with streams ingested}
  \label{fig:edge_results}
  \vspace{-0.1in}
\end{figure}
\subsubsection{Edge Analytics}

We evaluate the scalability of edge analytics on the testbed cluster of Nvidia Jetson devices by increasing the number of ingested RTSP streams. As expected, \textit{cumulative output FPS} processed scales with input load as long as no device exceeds its empirically determined FPS capacity (\S~\ref{sec:intelligent-routing}), ensuring real-time performance even with over \textit{2000 frames processed per second} (Fig.~\ref{fig:edge_results_1}).
To further quantify \textit{performance and power efficiency}, we track the \textit{Total Active Capacity} (sum of TOPS of active devices) and \textit{Utilization} (load relative to per-device FPS capacity), shown in Fig.~\ref{fig:edge_results_2}, alongside \textit{total power draw}. We compare the two stream-assignment strategies.

\textit{Best Fit} (upper plots) tightly packs streams onto the smallest suitable device, saturating the 32~GB Orins before activating the 64~GB ones. This produces a stepwise activation pattern, minimizes the number of active devices at moderate workloads, and reduces power consumption by deferring high-capacity nodes.
\textit{Worst Fit} (lower plots) distributes load across devices with the most remaining capacity, engaging larger devices earlier and improving load balance. However, this activates more devices sooner, increasing baseline power use. That said, in a heterogeneous cluster, the effect is more nuanced: Worst Fit may preferentially load high-capacity devices with better power-per-stream efficiency, \textit{sometimes yielding comparable or even lower power than Best Fit}. E.g., at 32 streams, Worst Fit consumes 231.6W versus 249.6W for Best Fit. These results show that both policies maintain real-time throughput, with Best Fit optimizing for minimal activation and Worst Fit improving balance.
and demonstrating effective scale-out across heterogeneous edge accelerators.

\begin{figure}[!t]
\vspace{-0.2in}
  \centering
    \subfloat[Vehicle Counts\label{subfig:ingest-1}]{
        \includegraphics[width=0.5\linewidth]{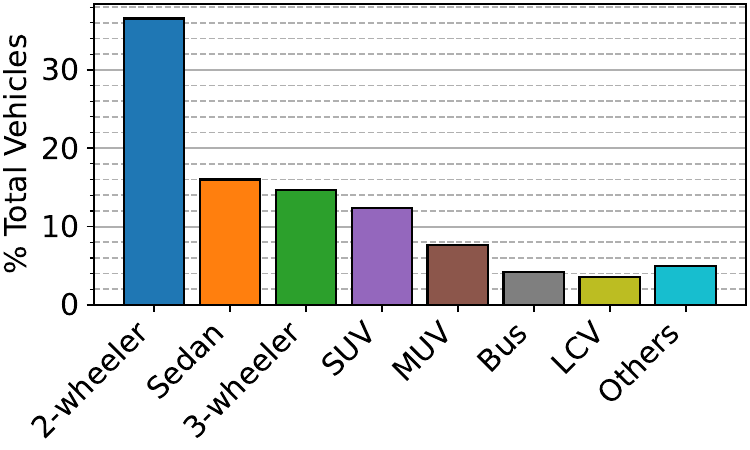}
    }
    \subfloat[Cumm. Vehicles/sec.~(100 stm.)\label{subfig:ingest-2}]{
        \includegraphics[width=0.45\linewidth]{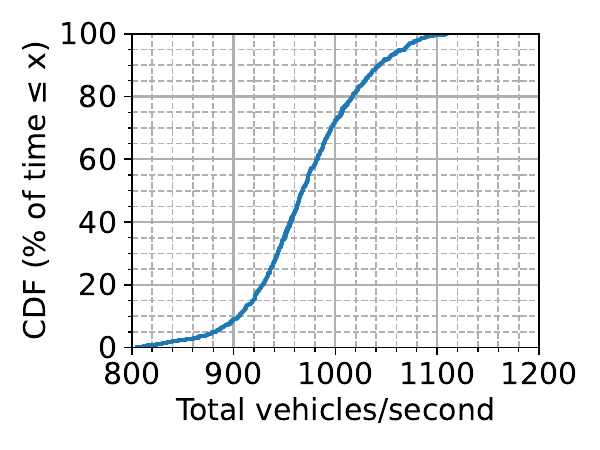}
    }\\
    \subfloat[GNN Train Perf.\label{subfig:gnn-1}]{
        \includegraphics[width=0.32\linewidth]{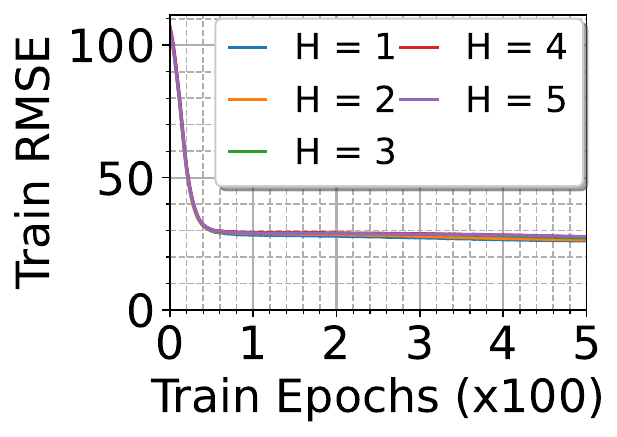}
    }
    \subfloat[GNN Deployment\label{subfig:gnn-2}]{
        \includegraphics[width=0.33\linewidth]{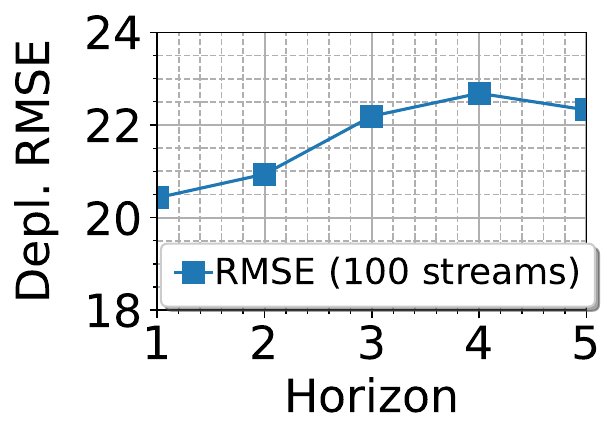}
    }
    \subfloat[Forecast Scaling\label{subfig:gnn-3}]{
        \includegraphics[width=0.32\linewidth]{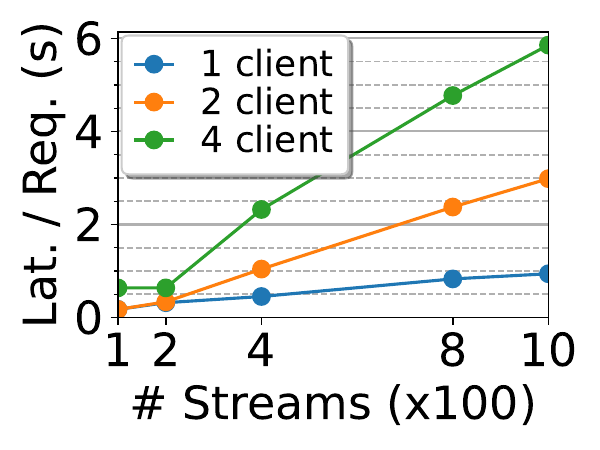}
    }
	\caption{Ingest Rate and GNN Performance}
    \label{fig:results:gnn}
\end{figure}

\begin{figure}[!t]
  \centering
    \includegraphics[width=0.9\columnwidth]{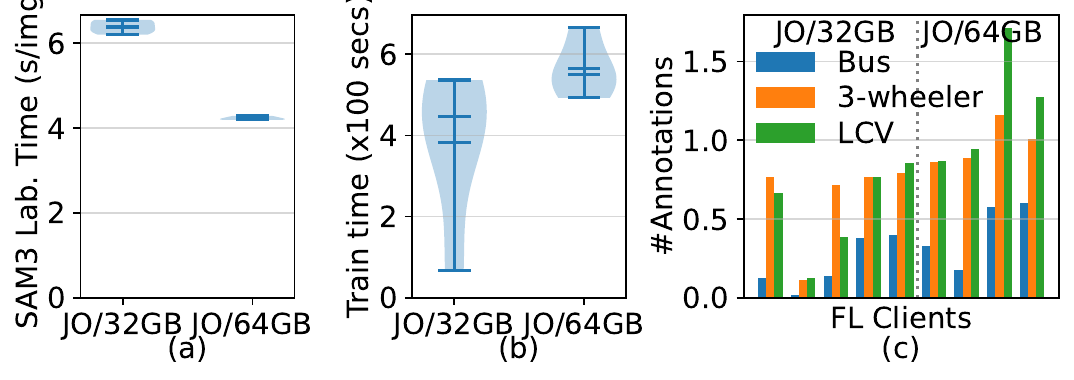}
	\caption{Performance of SAM3 labeling and FL training.}
    \label{fig:results:fl}
    \vspace{-0.1in}
\end{figure}

\subsection{Scalability of Ingest and ST-GNN Forecasting}

We analyze vehicle-flow data from the ingest service alongside the ST-GNN forecasting pipeline. Across 100 streams over 15 minutes, the system captures a diverse vehicle mix, with two-wheelers (37\%), sedans (15\%), and three-wheelers (14\%) forming the majority (Fig.~\ref{subfig:ingest-1}). Aggregate detections exceed 1000 unique vehicles/s for nearly 30\% of the window, peaking at 1110 vehicles/s, directly reflecting ingest throughput and exhibiting scalability under sustained load (Fig.~\ref{subfig:ingest-2}). TrendGCN training converges quickly across all horizons, with RMSE stabilizing early (Fig.~\ref{subfig:gnn-1}). In deployment, RMSE increases modestly from 20 at a 1-minute horizon to approximately 23 at 4 minutes and beyond, indicating stable multi-step forecasting performance (Fig.~\ref{subfig:gnn-2}). To assess scalability, we increase stream count from 100 to 1000 (augmenting with synthetic streams). With forecasts generated every 5 seconds, latencies remain manageable as concurrent clients scale from 1 to 4 (Fig.~\ref{subfig:gnn-3}), confirming that the ST-GNN pipeline sustains both higher camera volumes and client demand.

\subsection{Scalability of FM and Continuous FL}
To evaluate the practicality and scalability of the FL framework, we deploy the system on a cluster of 9 \emph{Nvidia Jetson Orin AGX}, $5\times32$GB~(\textsf{JO/32GB}) and $4\times64$GB~(\textsf{JO/64GB}). These Jetson devices collectively ingest 300 concurrent video streams, with each \textsf{JO/32GB} hosting 28 streams and \textsf{JO/64GB} hosting  40 streams. As detailed in \S\ref{subsec:fl}, we employ a temporally stratified sampling to extract one frame randomly every 20-second window, over 150 minutes, accruing 45 frames per video stream, totaling 1260 frames total for each \textsf{JO/32GB} and 1800 frames total for each \textsf{JO/64GB}. These frames are labeled by SAM3 leveraging the Orin's accelerators, with average per-image annotation latencies being 6.3s (\textsf{JO/32GB}) and 4.0s (\textsf{JO/64GB}), as seen in Fig.~\ref{fig:results:fl}(left). For each of the 9 FL clients, we show the distribution of the three classes unknown to the YOLO model in Fig.\ref{fig:results:fl}~(right) and the corresponding train time distribution for each Jetson type is reported in Fig.\ref{fig:results:fl}~(center). Since the \textsf{JO/64GB} receives a larger number of data streams, it collects more data, collecting about $1.2$--$5\times$ the data collected by the \textsf{JO/32GB}, processing more data per epoch, with a marginally higher training time. The annotation distribution shows non-IIDness of the distribution, highlighting the need for FL in this scenario to ensure all Jetsons are able to recognize all classes.

\subsection{Discussion and Scaling Demo}

Our AIITS architecture achieves extensibility by functionally isolating edge-inference from cloud-based spatio-temporal forecasting. This facilitates horizontal scale-out of accelerators without systemic reconfiguration. Our approach leverages a novel capacity-aware elastic scheduler and SAM3-driven labeling to mitigate ``data drift'' expected within unstructured urban environments. By localizing heavy video analytics at the network periphery, the system avoids centralized bandwidth bottlenecks, enabling sustainable, city-scale traffic sensing.

The SCALE Challenge live demo will exhibit the full scalability of our system by expanding beyond the paper's results to support 1000 video streams across $\approx100$ RPis and $\approx 25$ Jetson accelerators.
This will highlight the ability of our design to sustain city-scale, heterogeneous, multi-modal workloads in real time.

\section{Related Work}

Intelligent Transportation Systems (ITS) often rely on dedicated roadway sensors (e.g., inductive loops) that directly provide usable time-series data but are costly to scale city-wide in emerging nations~\cite{zhou2021intelligent}. While CCTV networks are widespread, converting large volumes of video feeds into traffic-flow signals is less explored and requires distributed multi-stream detection, tracking, synchronization, and aggregation~\cite{barthelemy2019edge,yang2023cooperative}. Prior systems address individual components rather than providing an end-to-end pipeline with an integrated solution, as we do. Li et al.~\cite{li2025multi} propose a multi-stage surveillance framework but with limited scaling; Seo et al.~\cite{seo2025dnnpipe} optimize DNN partitioning but not concurrent multi-stream workloads; while Rong et al.~\cite{rong2021scheduling} study edge--cloud scheduling under homogeneous assumptions. 

Spatio-temporal GNNs~\cite{TrendGCN,lan2022dstagnn,jiang2023pdformer} achieve strong forecasting performance but assume synchronized sensor inputs rather than video-derived signals produced at scale. A further gap lies in model adaptation. Pretrained detectors often underperform in heterogeneous and emerging city traffic, while constructing large annotated datasets is costly. Open-world discovery methods reduce manual labeling but typically assume centralized data access~\cite{Vaze2022-GCD}. FL enables local training on the edge while avoiding raw data sharing~\cite{fedavg}, yet device heterogeneity remains challenging~\cite{Li_2020}. Motivated by these gaps, we integrate heterogeneous multi-stream edge inference, capacity-aware scheduling, graph-level aggregation, GNN forecasting and SAM-assisted labeling with lightweight FL updates into a unified end-to-end analytics pipeline for city-scale predictions.

\section{Conclusions}
We presented an AIITS platform that combines lightweight DNNs, ST-GNN forecasting, foundation-model-assisted labeling, and continuous FL to scale efficiently across an edge--cloud fabric while minimizing data movement and centralized compute. The results confirm its extensibility across large camera deployments and heterogeneous edge hardware, with multi-modal scaling: stream-level scale-out, device-level scale-up, and cross-fabric coordination, demonstrating its practical impact on city-scale urban mobility analytics. 
In the future, we will evaluate performance under more congested conditions, extend the pipeline to 5000+ feeds, and study the performance interplay between inference and FL. We plan to further optimize stream to Jetson placement using network topology and energy signals, and enable dynamic model selection to sustain throughput with variable streams.

\section{Acknowledgments}
We thank the Bengaluru Traffic Police (BTP) and the Bengaluru Police for providing access to the Safe City camera data from which the video data used in this study were derived.

\bibliographystyle{plain}
\bibliography{references.bib}

@techreport{sharma2025uvh26,
  title        = {The Urban Vision Hackathon Dataset and Models: Towards Image Annotations and Accurate Vision Models for Indian Traffic},
  author       = {Sharma, Akash and Mhatre, Chinmay and Gawali, Sankalp and Bokkasam, Ruthvik and Kishore, Brij and Pattanaik, Vishwajeet and Rambha, Tarun and Pinjari, Abdul R. and Kovvali, Vijay and Chakraborty, Anirban and Rathore, Punit and Krishnapuram, Raghu and Simmhan, Yogesh},
  institution  = {Indian Institute of Science},
  OPTtype         = {Technical Report},
  number       = {arXiv:2511.02563},
  year         = {2025},
  OPTurl          = {https://doi.org/10.48550/arXiv.2511.02563},
  OPTnote         = {Version 1, submitted on 4 Nov 2025},
}

@inproceedings{li2018diffusion,
title={Diffusion Convolutional Recurrent Neural Network: Data-Driven Traffic Forecasting},
author={Yaguang Li and Rose Yu and Cyrus Shahabi and Yan Liu},
booktitle={International Conference on Learning Representations},
year={2018},
OPTurl={https://openreview.net/forum?id=SJiHXGWAZ},
}

@software{ultralytics_yolo,
  author = {Glenn Jocher and Jing Qiu},
  title = {Ultralytics YOLO},
  year = {2023},
  publisher = {GitHub},
  journal = {GitHub repository},
  howpublished = {\url{https://github.com/ultralytics/ultralytics}}
}

@techreport{aharon2022botsortrobustassociationsmultipedestrian,
    author={Nir Aharon and others} ,
    title={BoT-SORT: Robust Associations Multi-Pedestrian Tracking},
    year={2022},
    institution = {arXiv},
    number={2206.14651}
}

@incollection{coffman1984approximation,
  author    = {Coffman, Edward G. and Garey, Michael R. and Johnson, David S.},
  title     = {Approximation Algorithms for Bin-Packing --- An Updated Survey},
  booktitle = {Algorithm Design for Computer System Design},
  editor    = {Ausiello, Giorgio and Lucertini, Mario and Serafini, Paolo},
  series    = {International Centre for Mechanical Sciences},
  volume    = {284},
  publisher = {Springer},
  address   = {Vienna},
  year      = {1984},
  doi       = {10.1007/978-3-7091-4338-4_3}
}

@inproceedings{fedavg,
  author = {cMahan, H. Brendan and Moore, E. and Ramage, D. and Hampson, S. and y Arcas, B. A.},
  title = {Communication-Efficient Learning of Deep Networks from Decentralized Data},
  booktitle = {AISTATS},
  year = {2017}
}

@inproceedings{lan2022dstagnn,
  title={Dstagnn: Dynamic spatial-temporal aware graph neural network for traffic flow forecasting},
  author={Lan, Shiyong and Ma, Yitong and Huang, Weikang and Wang, Wenwu and Yang, Hongyu and Li, Pyang},
  booktitle={International conference on machine learning},
  pages={11906--11917},
  year={2022},
  organization={PMLR}
}

@inproceedings{jiang2023pdformer,
  title={Pdformer: Propagation delay-aware dynamic long-range transformer for traffic flow prediction},
  author={Jiang, Jiawei and Han, Chengkai and Zhao, Wayne Xin and Wang, Jingyuan},
  booktitle={Proceedings of the AAAI conference on artificial intelligence},
  volume={37},
  number={4},
  pages={4365--4373},
  year={2023}
}

@inproceedings{TrendGCN,
author = {Jiang, Juyong and Wu, Binqing and Chen, Ling and Zhang, Kai and Kim, Sunghun},
title = {Enhancing the Robustness via Adversarial Learning and Joint Spatial-Temporal Embeddings in Traffic Forecasting},
year = {2023},
isbn = {9798400701245},
publisher = {Association for Computing Machinery},
address = {New York, NY, USA},
url = {https://doi.org/10.1145/3583780.3614868},
doi = {10.1145/3583780.3614868},
booktitle = {Proceedings of the 32nd ACM International Conference on Information and Knowledge Management},
pages = {987–996},
numpages = {10},
location = {Birmingham, United Kingdom},
series = {CIKM '23}
}

@inproceedings{coco,
  title={Microsoft coco: Common objects in context},
  author={Lin, Tsung-Yi and Maire, Michael and Belongie, Serge and Hays, James and Perona, Pietro and Ramanan, Deva and Doll{\'a}r, Piotr and Zitnick, C Lawrence},
  booktitle={European conference on computer vision},
  pages={740--755},
  year={2014},
  organization={Springer}
}

@article{naing2024fine,
  title={Fine-grained trajectory reconstruction by microscopic traffic simulation with dynamic data-driven evolutionary optimization},
  author={Naing, Htet and Cai, Wentong and Yu, Jinqiang and Zhong, Jinghui and Yu, Liang},
  journal={IEEE Transactions on Intelligent Transportation Systems},
  volume={26},
  number={2},
  pages={1930--1950},
  year={2024},
  publisher={IEEE}
}

@article{rong2021scheduling,
  title={Scheduling massive camera streams to optimize large-scale live video analytics},
  author={Rong, Chenghao and Wang, Jessie Hui and Liu, Juncai and Wang, Jilong and Li, Fenghua and Huang, Xiaolei},
  journal={IEEE/ACM Transactions on Networking},
  volume={30},
  number={2},
  pages={867--880},
  year={2021},
  publisher={IEEE}
}

@article{li2025multi,
  title={A multi stage deep learning approach for real-time vehicle detection, tracking, and speed measurement in intelligent transportation systems},
  author={Li, Ran},
  journal={Scientific reports},
  volume={15},
  number={1},
  pages={22531},
  year={2025},
  publisher={Nature Publishing Group UK London}
}

@inproceedings{Vaze2022-GCD,
  author = {Vaze, S. and Balasubramanian, V. and Gygli, M. and Nestarev, Y. and Vedaldi, A.},
  title = {Generalized Category Discovery},
  booktitle = {CVPR},
  year = {2022},
}

@article{seo2025dnnpipe,
  title={DNNPipe: Dynamic programming-based optimal DNN partitioning for pipelined inference on IoT networks},
  author={Seo, Woobean and Kim, Saehwa and Hong, Seongsoo},
  journal={Journal of Systems Architecture},
  volume={166},
  pages={103462},
  year={2025},
  publisher={Elsevier}
}

@article{barthelemy2019edge,
  title={Edge-computing video analytics for real-time traffic monitoring in a smart city},
  author={Barth{\'e}lemy, Johan and Verstaevel, Nicolas and Forehead, Hugh and Perez, Pascal},
  journal={Sensors},
  volume={19},
  number={9},
  pages={2048},
  year={2019},
  publisher={MDPI}
}

@article{yang2023cooperative,
  title={Cooperative multi-camera vehicle tracking and traffic surveillance with edge artificial intelligence and representation learning},
  author={Yang, Hao Frank and Cai, Jiarui and Liu, Chenxi and Ke, Ruimin and Wang, Yinhai},
  journal={Transportation research part C: emerging technologies},
  volume={148},
  pages={103982},
  year={2023},
  publisher={Elsevier}
}

@article{zhou2021intelligent,
  title={When intelligent transportation systems sensing meets edge computing: Vision and challenges},
  author={Zhou, Xuan and Ke, Ruimin and Yang, Hao and Liu, Chenxi},
  journal={Applied Sciences},
  volume={11},
  number={20},
  pages={9680},
  year={2021},
  publisher={MDPI}
}

@misc{tomtom-2025,
    author = {{TomTom}},
    title = {{TomTom} Traffic Index Ranking 2025},
    year = {2025},
    howpublished = {\url{https://www.tomtom.com/traffic-index/ranking}},
    note = {Accessed: 2026-02-19}
}

@misc{iea_energy_efficiency_2025_transport,
    author = {{International Energy Agency (IEA)}},
    title = {Energy Efficiency 2025: Transport},
    year         = {2025},
    publisher    = {International Energy Agency},
    howpublished = {\url{https://www.iea.org/reports/energy-efficiency-2025/transport}},
    note         = {Accessed: Feb. 19, 2026}
}

@techreport{efd_dp_23_10_congestion_health_hazards,
  author       = {Kacker, Kanishka and Gupta, Ridhima and Ali, Saif},
  title        = {Does Traffic Congestion pose Health Hazards? Evidence from a Highly Congested and Polluted City},
  institution  = {Environment for Development (EfD) Initiative},
  year         = {2023},
  month        = jul,
  type         = {Discussion Paper Series},
  number       = {DP 23-10},
  howpublished = {\url{https://www.efdinitiative.org/sites/default/files/publications/EfD_DP-23-10.pdf}},
  note         = {Accessed: Feb. 19, 2026}
}

@techreport{grsf_annual_report_2024,
    author = {{Global Road Safety Facility (GRSF)}},
    title = {{GRSF} Annual Report 2024},
    institution = {The Global Road Safety Facility, World Bank},
    year = {2024},
    howpublished = {\url{https://www.globalroadsafetyfacility.org/sites/default/files/2024-11/GRSF%20Annual%20Report%202024_1.pdf}},
    note         = {Accessed: Feb. 19, 2026}
}

@article{Li_2020,
   title={Federated Learning: Challenges, Methods, and Future Directions},
   OPTvolume={37},
   OPTISSN={1558-0792},
   OPTDOI={10.1109/msp.2020.2975749},
   OPTnumber={3},
   journal={IEEE Signal Processing Magazine},
   OPTpublisher={Institute of Electrical and Electronics Engineers (IEEE)},
   author={Li, Tian and Sahu, Anit Kumar and Talwalkar, Ameet and Smith, Virginia},
   year={2020},
   OPTmonth=may, OPTpages={50–60} 
}

@article{carion2025sam,
  title={Sam 3: Segment anything with concepts},
  author={Carion, Nicolas and Gustafson, Laura and Hu, Yuan-Ting and Debnath, Shoubhik and Hu, Ronghang and Suris, Didac and Ryali, Chaitanya and Alwala, Kalyan Vasudev and Khedr, Haitham and Huang, Andrew and others},
  journal={arXiv preprint arXiv:2511.16719},
  year={2025}
}

\end{document}